\documentclass[12pt]{iopart}

\usepackage{hyperref}
\usepackage{graphicx,epsfig,dsfont,amsthm}
\begin{document}
\eqnobysec

\newtheorem{thm}{Theorem}
\newtheorem{lem}{Lemma}
\newtheorem{defi}{Definition}
\newcommand{\ren}[3]{S_{#1}^{#2}\left(#3\right)}
\newcommand{\hmin}[3]{H_{\mathrm{min}}^{#1}\left(#2|#3\right)}
\newcommand{\hmax}[3]{H_{\mathrm{max,old}}^{#1}\left(#2|#3\right)}
\newcommand{\hmaxnew}[3]{H_{\mathrm{max,new}}^{#1}\left(#2|#3\right)}
\newcommand{\leak}{\mathrm{leak}_{\mathrm{EC}}} 
\newcommand{\R}{\mathbb{R}}
\newcommand{\ball}[2]{{\cal B}^{#1}\left(#2\right)}
\newcommand{\pr}[1]{{\ketbra{#1}{#1}}}
\newcommand{\trz}[2]{\mathrm{tr}_{#1}\left( #2\right)}
\newcommand{\epsbarec}{\bar{\varepsilon}_{\mathrm{EC}}}
\newcommand{\hnull}[3]{H_{0}^{#1}\left(#2|#3\right)}

\newcommand{\bra}[1]{\left\langle{#1}\right\vert}
\newcommand{\ket}[1]{\left\vert{#1}\right\rangle}
\newcommand{\scalar}[2]{\left\vert \left\langle {#1}\vert {#2} \right\rangle \right\vert^2}
\newcommand{\norm}[1]{\bigl\|{#1}\bigr\|_1}
\newcommand{\epsbar}{\bar{\varepsilon}}
\newcommand{\epspa}{\varepsilon_{\mathrm{PA}}}
\newcommand{\overlap}[2]{\left<{#1}|{#2}\right>}
\newcommand{\proj}[1]{\left\vert{#1}\right\rangle \left\langle{#1}\right\vert}
\newcommand{\p}[1]{P_{\left\vert{#1}\right\rangle}}
\newcommand{\Prob}[1]{\mathrm{Prob}\left[#1\right]}
\newcommand{\dist}[1]{\frac{1}{2}\left|\left|#1\right|\right|_1} 
\newcommand{\bracket}[2]{{\langle #1 | #2 \rangle}}
\newcommand{\braccket}[3]{{\langle #1 | #2 | #3 \rangle}}
\newcommand{\ketbra}[2]{{|#1\rangle\!\langle#2|}}
\newcommand{\id}{{1\!\!1}}
\newcommand{\eps}[1]{\varepsilon_{\mathrm{#1}}}

\title{Quantum key distribution with finite resources: Taking advantage of quantum noise}

\author{M. Mertz, H. Kampermann, Z. Shadman and D. Bru{\ss}}
\address{Institute for Theoretical Physics III, Heinrich-Heine-Universit\"at D\"usseldorf, 40225 D\"usseldorf, Germany.}
\ead{mertz@thphy.uni-duesseldorf.de}

\date{\today}

\begin{abstract}
We compare the effect of different  noise scenarios on the achievable rate of an $\eps{}$-secure key for the BB$84$ and the six-state protocol. We study the situation where quantum noise is added deliberately, and investigate the remarkable benefit for the finite key rate. We compare our results to the known case of added classical noise and the asymptotic key rate, i.e. in the limit of infinitely many signals. As a complementary interpretation we show that under the realistic assumption that the noise which is unavoidably introduced by a real channel is not fully 
dedicated to the eavesdropper, the secret key rate increases significantly.
\end{abstract}

\maketitle

\section{Introduction}
Quantum key distribution (QKD) aims at  establishing  a secret key between two parties Alice and Bob, who are connected via a quantum channel and an authenticated classical channel. In the last few years, in addition to the studies of asymptotic QKD (i.e. the unrealistic case
of infinitely many signals, which is more accessible theoretically), more realistic QKD scenarios have been analyzed, where the number of signals sent through the channel is finite \cite{RenPhD,RenKoe,Ren05,Mey06,Sca08,Sca08a,Sca09,Cai09,She10,Bra11,Tom11,Abr11,Hay11}.

A general aim in  studies of security in QKD is to determine a scenario in which the secure key rate is
as high as possible.
It has been shown that pre-processing methods \cite{RenPhD,Bae07,Renes07,Smi08,Kern08,Rene10,Rene10a}, like for example adding classical noise \cite{RenPhD} or advantage distillation \cite{RenPhD} can increase the secure key rate significantly. Note that  those investigations have focused on pre-processing operating on the {\em classical} level. By the addition of {\em quantum noise} a beneficial effect in asymptotic QKD (on the level of mutual information) has been shown in \cite{Shad09} for the six-state protocol \cite{Bru98,Bech99}. 
However, for a finite number of signals the mutual information of Alice and Bob versus the one of the eavesdropper Eve
is not a direct indicator for the secret key rate.
 
The purpose of this article is to investigate the effect of {\em quantum} noise on secret key rates with {\em finite} resources for the BB$84$ \cite{BB84} and six-state protocol \cite{Bru98,Bech99}. We will analyse our results for two complementary interpretations: First, we present different quantum noise scenarios, where the noise is added on purpose, and investigate its benefit for the secret key rate. Second, we interpret the added noise as the unavoidable noise introduced by a real channel. We then show how the secret key rate can be improved if we consider the noise introduced by the channel as not fully due to the interaction of  an eavesdropper. We compare the results to the known effect of classical noise and the case of infinitely many signals. For the investigation we consider the BB$84$ and six-state protocol in the entanglement-based scheme under the assumption of collective attacks. We use the asymptotic equipartition property (AEP) \cite{RenPhD,Tom09} to bound the smooth min-entropy \cite{RenPhD} in the high-dimensional Hilbert space, such that the $\eps{}$-secure key rate can be mainly determined by the conditional von Neumann entropy of a single-signal-state. Note that here, for key rates in the finite regime, the assumption of collective attacks is necessary, since the equivalence of collective and coherent attacks for the BB$84$ and the six-state protocol has so far been proven only in the limit of infinitely many signals \cite{Kra05,Ren05a}.

The paper is structured as follows: In \Sref{SEC:preliminaries} we introduce the general framework and fix the notation. The different noise scenarios are presented and discussed in \Sref{SEC:noisescenarios}. \Sref{SEC:keyrate} deals with the calculation and optimization of $\eps{}$-secure key rates for these different noise scenarios. The results are given in \Sref{SEC:results}, followed by a conclusion in \Sref{SEC:conclusion}.

\section{Preliminaries}\label{SEC:preliminaries}
In the following we consider the BB$84$ and six-state protocol in the entanglement-based scheme, where the eavesdropper Eve can only interact with the signals (labeled by $B$) which are sent through the quantum channel. The most general unitary interaction $U_{BE}$ that Eve can perform is given by \cite{Fuchs96}
\begin{eqnarray}\label{EQ:interact}
 U_{BE}\ket{0}_B\ket{X}_E&=&\sqrt{1-D}\ket{0}_B\ket{A}_E+\sqrt{D}\ket{1}_B\ket{B}_E \\
 U_{BE}\ket{1}_B\ket{X}_E&=&\sqrt{1-D}\ket{1}_B\ket{C}_E+\sqrt{D}\ket{0}_B\ket{D}_E,
\end{eqnarray}
 where $\ket{X}_E$ is Eve's initial state and $\ket{A}_E$, $\ket{B}_E$, $\ket{C}_E$, $\ket{D}_E$ refer to her $4$-dimensional states after the transformation. 
The parameter $D \in \left[0,\frac{1}{2}\right]$ corresponds to the disturbance, i.e. the quantum bit error rate ($QBER$) introduced by Eve if the quantum channel is otherwise noiseless. 

Throughout our paper we will study quantum noise which is given by a depolarizing channel.
(Note that our calculations could in principle be generalized to other models for quantum noise, but the
lower the symmetry of the channel, the more involved the calculations will be.) The action of the depolarizing channel is
  described by the map $\mathcal{N}^p\left( \rho \right)$, where $p$ is the  noise parameter.
\begin{defi}\label{defi:deponoise} The action of a depolarizing channel $\mathcal{N}^p\left( \rho \right)$ is given by
\begin{equation}
 \mathcal{N}^p\left( \rho \right):=\sum_{i=1}^4 A_i \rho A^\dagger_i
\end{equation}
with the Kraus operators $A_1=\sqrt{1-\frac{3}{4}p}\id, A_2=\sqrt{\frac{p}{4}}\sigma_x, A_3=\sqrt{\frac{p}{4}}\sigma_y, 
and A_4=\sqrt{\frac{p}{4}}\sigma_z$. Here, $\sigma_i$ are the Pauli-operators for $i\in \left\lbrace x,y,z\right\rbrace $.
\end{defi} Analogously, we define classical noise \cite{RenPhD} via the map $\mathcal{N}^{cl,p}\left( \rho \right)$. 
\begin{defi}\label{defi:classnoise} The action of a classical noisy channel is given by
\begin{equation}
 \mathcal{N}^{cl,p}\left( \rho \right):=\sum_{i=1}^2 B_i \rho B^\dagger_i
\end{equation} with $B_1=\sqrt{1-\frac{p}{2}}\id$ and $B_2=\sqrt{\frac{p}{2}} \sigma_x$.
\end{defi} Note that this definition is different from the usual definition of classical noise in the literature: throughout our paper the
 probability to flip a bit is called $\frac{p}{2}$ instead of $p$. This choice of $p$ allows a fair comparison of the two different noise models 
(quantum versus classical) for the same parameter $p$, ranging from $0$ to $1$.

Our central figure of merit is the $\eps{}$-secure key rate for a finite number of signals. We will
use this quantity in the following 
to compare different noise scenarios.
The $\eps{}$-secure key rate is calculated for a typical protocol that consists of the procedures state distribution, measurement, sifting, parameter estimation (PE), one-way error correction (EC) and privacy amplification (PA). Let $\eps{PE}, \eps{EC}$ and $\eps{PA}$ be the probability of failure for the protocol steps parameter estimation, error correction and privacy amplification, respectively. Then with a smoothing parameter $\bar{\varepsilon}$ we can bound the total security of the protocol by 
\begin{equation}
\varepsilon:=\bar{\varepsilon}+\varepsilon_{\mathrm{PE}}+\varepsilon_{\mathrm{EC}}+\varepsilon_{\mathrm{PA}}.
\end{equation}
For such a protocol it has been shown in \cite{RenPhD,Sca08a} that the rate of an $\varepsilon$-secure key is given by 
\begin{equation}
r=\frac{1}{N}\min_{\rho_{AB}\in\Gamma_\zeta} \left(\hmin{\bar{\varepsilon}}{\rho^n_{XE}}{E}-nf_{\mathrm{EC}}H(X|Y)\right)+\frac{2}{N}\log\left(2\varepsilon_\mathrm{PA}\right),
\end{equation} 
where the smooth min-entropy \cite{RenPhD}
\begin{equation}
 \hmin{\eps{}}{\rho_{AE}}{E}:= \sup_{\sigma_{AE} \in \ball{\frac{\eps{}}{2}}{\rho_{AE}}} \sup_{\rho_E \in S(\mathcal{H})} \hmin{}{\sigma_{AE}}{\rho_E}
\end{equation}
is defined as an optimization of the min-entropy
\begin{equation}
   \hmin{}{\sigma_{AE}}{\rho_E}:=\sup{\left\lbrace \lambda \in \Re:2^{-\lambda}\id_A\otimes \rho_E-\sigma_{AE}\geq0\right\rbrace }
 \end{equation} 
over an $\eps{}$-environment given by 
\begin{equation}
  \ball{\eps{}}{\rho}:=\left\lbrace \sigma: \dist{\sigma -\rho}\leq \eps{} \right\rbrace,
\end{equation}
with the $1$-norm $\left|\left|A\right|\right|_1=\tr\left(\sqrt{AA^\dagger}\right)$. Here, $S(\mathcal{H})$ denotes the set of density operators on the Hilbert space $\mathcal{H}$.
The smooth min-entropy of the classical-quantum state $\rho^n_{XE}$ shared by Alice and Eve and the correction $2\log_2\left(2\eps{PA}\right)$ are due to privacy amplification. It quantifies Eve's uncertainty of Alice's and Bob's perfectly correlated bitstring. 
The term $f_{\mathrm{EC}}H(X|Y)$ stands for the number of bits which Alice and Bob leak to the eavesdropper due to public communication during the error correction procedure. $H(X|Y)$ denotes the conditional Shannon entropy $H(X|Y)=H(\rho_{XY})-H(\rho_Y)$ with $H(X)=-\sum_x p(x)\log{\left(p(x)\right)}$. For simplicity we consider an ideal error correction protocol, i.e. $f_{\mathrm{EC}}=1$.
The minimization of the smooth min-entropy is due to parameter estimation, where we only except qubit-states $\rho_{AB}$ which are contained in the set \cite{Sca08,Bra11} 
\begin{equation}\label{EQ:PE}
 \Gamma_\zeta:=\left\lbrace \rho: \dist{\lambda_m(\rho)-\lambda_\infty(\rho)}\leq \zeta(\eps{PE},2,m) \right\rbrace 
\end{equation} with
\begin{equation}
\zeta(\eps{PE},n_p,m):=\sqrt{\frac{\ln{\left(\frac{1}{\eps{PE}}\right)}+n_p\ln{(m+1)}}{8m}},
\end{equation} where $\lambda_m(\rho)$ ($\lambda_\infty(\rho)$) denotes the measurement statistics due to an $m$ ($m\rightarrow \infty$)-fold independent application of a measurement.

Under the assumption of collective attacks, i.e. $\rho^n_{XE}=\rho_{XE}^{\otimes n}$ we can use the AEP \cite{RenPhD,Tom09}
\begin{equation}
\hmin{\eps{}}{\rho^{\otimes n}_{XE}}{E}\geq n\left(S(X|E)-5\sqrt{\frac{\log{\left(2/\eps{}\right)}}{n}}\right)
\end{equation} to bound the smooth min-entropy for product states $\rho_{XE}^{\otimes n}$ by the conditional von Neumann entropy of a single copy $\rho_{XE}$ which is defined as $S(X|E)=S(\rho_{XE})-S(\rho_E)$ with $S(\rho)=-\tr{\left(\rho \log{\rho}\right)}$.
Finally, this leads to
\begin{eqnarray}\label{EQ:finitekey}
r&:=&\frac{n}{N}\min_{\rho_{AB}\in\Gamma_\zeta} \left(S(X|E)-5\sqrt{\frac{\log(2/\bar{\varepsilon})}{n}}-H(X|Y)\right) 
+\frac{2}{N}\log\left(2\varepsilon_\mathrm{PA}\right).
\end{eqnarray} 

\section{Noise scenarios}\label{SEC:noisescenarios}
In this section we present four different noise scenarios which we will investigate in the following. Initially, if no noise is present, Alice holds one part of the Bell-state $\ket{\Psi^+}\bra{\Psi^+}$, with 
\begin{equation}\label{EQ:MaxEnt}
\ket{\Psi^+}=\frac{1}{\sqrt{2}}\left(\ket{01}+\ket{10}\right),
\end{equation} and sends the second part to Bob, while the eavesdropper can perform a unitary interaction $U_{BE}$ characterized by a disturbance $D$ (see \Eref{EQ:interact}) on it. Let us denote the map that corresponds to the unitary interaction $U_{BE}$ as $\mathcal{E}_{BE}$. The total state after the action of $\mathcal{E}_{BE}$ is then given by
\begin{equation} 
\rho^{(0)}_{ABE}=(\id_A \otimes \mathcal{E}_{BE})\left(\ket{\Psi^+}\bra{\Psi^+}_{AB}\otimes \proj{X}_E \right), 
\end{equation} where $\proj{X}_E$ denotes Eve's initial state. 

\begin{figure}
\centerline{\includegraphics[width=0.5\textwidth]{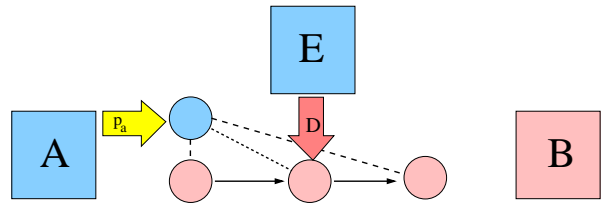}}
\caption{Noise scenario $1$; 
Alice adds depolarizing quantum noise to her part of the initial state.
}
\label{Fig:noisescenpa}
\end{figure}

\begin{figure}
\centerline{\includegraphics[width=0.5\textwidth]{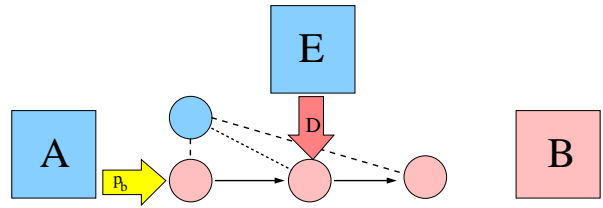}}
\caption{Noise scenario $2$; 
Alice adds depolarizing quantum noise to Bob's part of the initial state.
}
\label{Fig:noisescenpb}
\end{figure}

\begin{figure}
\centerline{\includegraphics[width=0.5\textwidth]{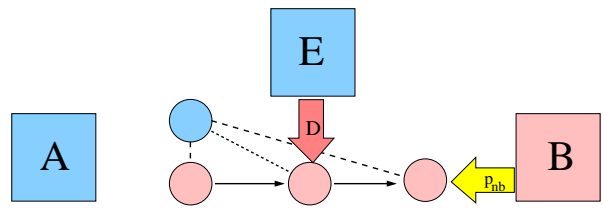}}
\caption{Noise scenario $3$; 
Bob adds depolarizing quantum noise to his part of the state after receiving it.
}
\label{Fig:noisescenpnb}
\end{figure}

\begin{figure}
\centerline{\includegraphics[width=0.15\textwidth,angle=90]{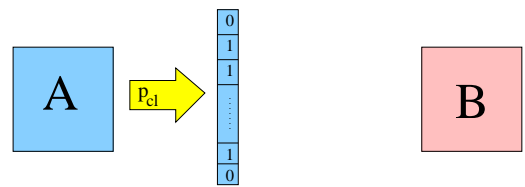}}
\caption{Noise scenario $4$; 
Alice adds classical noise to her classical bit string.
}
\label{Fig:noisescenpcl}
\end{figure}

Now, four different noise scenarios are considered:
\begin{enumerate}
 \item[(1)] Alice adds depolarizing quantum noise with noise parameter $p_a$ to her part of the Bell-state and sends the other part to Bob (see \Fref{Fig:noisescenpa}). This leads to the state  
 \begin{eqnarray}
\hspace*{-0.8cm}
\label{rhoabc}
 \rho^{(1)}_{ABE} &=& (\id_A \otimes \mathcal{E}_{BE})(\mathcal{N}^{p_a}_A \otimes \id_{BE}) 
            \left(\ket{\Psi^+}\bra{\Psi^+}_{AB}\otimes \proj{X}_E \right).
 \end{eqnarray} Note that there is obviously no difference between adding Alice's noise before or after Eve's interaction, since they act on different Hilbert spaces.
 \item[(2)] Alice adds depolarizing noise with noise parameter $p_b$ to Bob's part of the Bell-state and sends it to Bob (see \Fref{Fig:noisescenpb}). This leads to the state 
 \begin{eqnarray}
\hspace*{-0.8cm}
 \rho^{(2)}_{ABE} &=& (\id_A \otimes \mathcal{E}_{BE})(\id_A \otimes \mathcal{N}^{p_b}_B\otimes \id_E) 
                        \left(\ket{\Psi^+}\bra{\Psi^+}_{AB}\otimes \proj{X}_E \right).
 \end{eqnarray}
 \item[(3)] Bob adds depolarizing noise with noise parameter $p_{nb}$ to his part of the Bell-state after Eve's interaction (see \Fref{Fig:noisescenpnb}). This leads to the state 
 \begin{eqnarray}
\hspace*{-0.8cm}
 \rho^{(3)}_{ABE} &=& (\id_A \otimes \mathcal{N}^{p_{nb}}_B \otimes \id_E)(\id_A \otimes \mathcal{E}_{BE}) 
                    \left(\ket{\Psi^+}\bra{\Psi^+}_{AB}\otimes \proj{X}_E\right).
 \end{eqnarray}
 \item[(4)] Alice introduces classical noise with noise parameter $p_{cl}$ to her classical bit string after her measurement (see \Fref{Fig:noisescenpcl}).
\end{enumerate} 
How do these four scenarios compare, when evaluating the $\eps{}$-secure key rate?

\section{Secret key rate}\label{SEC:keyrate}
The aim of this section is to investigate the effect of the various noise scenarios explained in the previous section on the finite $\eps{}$-secure key rate $r$ in \Eref{EQ:finitekey}. 

From the fact that Alice and Bob share in the beginning a maximally entangled state (see \Eref{EQ:MaxEnt}) we know that the action of a depolarizing channel on Alice's part or on Bob's part results in the same total state. This implies that the states $\rho^{(1)}_{ABE}$ and $\rho^{(2)}_{ABE}$ are identical for $p_a=p_b$ and the noise scenarios $1$ and $2$ are equivalent.

Additionally, we now show the equivalence of noise scenario $1$ and $4$, i.e. adding rotationally invariant quantum noise is
equivalent to adding classical noise. Let us assume for noise scenario $4$ that we add classical noise 
 with probability $\frac{p_{cl}}{2}$ to flip a bit (see Definition~\ref{defi:classnoise})
to a bit-string resulting from measurements in the $z$-basis. In noise scenario $1$ only the Pauli-operators $\sigma_x$ and $\sigma_y$ from the depolarizing channel (see Definition~\ref{defi:deponoise}) lead to a bit-flip, such that the total probability to flip a bit is given by $\frac{p_a}{4}+\frac{p_a}{4}=\frac{p_a}{2}$. Note that for the cases in scenario $4$ that the bit-strings were obtained by measurements in $x$ ($y$)-basis the same argument holds. Then only $\sigma_y$ and $\sigma_z$ ($\sigma_x$ and $\sigma_z$) lead to bit-flips in scenario $1$, and due to the symmetry of the depolarizing channel the probability to flip is also $\frac{p_a}{2}$ in both cases. This implies the equivalence of scenario $1$ and scenario $4$ for $p_a=p_{cl}$.

Noise scenario $3$ will not lead to any benefit for the key rate as it only increases the quantum bit error rate ($QBER$). This asymmetry 
between noise scenario $2$ and $3$ is due to the underlying one-way error-correction protocol, such that adding Bob's noise after Eve's interaction only influences the key rate $r$ by increasing $H(X|Y)$.

The equivalence of the noise scenarios 1, 2 and 4, together with the fact that scenario 3 can only be
detrimental for the secret key rate, enable us to focus for the rest of this paper on a specific noise scenario, 
namely noise scenario $1$. We start for simplicity with the investigation of the asymptotic key rate (i.e. \Eref{EQ:finitekey} for $N\rightarrow \infty, \eps{}\rightarrow 0$) given by \cite{Dev05}
\begin{equation}\label{EQ:asymrate}
 r_{\mathrm{asym}}=S(X|E)-H(X|Y).
\end{equation}
Later,  the effect of noise  on the finite key rate (see \Eref{EQ:finitekey}) follows by including the finite-size effects.

In order to determine Eve's unknown probes in the state $\rho^{(1)}_{ABE}$, 
given in Eq. (\ref{rhoabc}), 
namely $\ket{A}_E, \ket{B}_E, \ket{C}_E, \ket{D}_E$, we expand each probe in basis vectors
\begin{equation} 
 \ket{A}_E=\alpha_a\ket{00}+\beta_a\ket{01}+\gamma_a\ket{10}+\delta_a\ket{11} 
\end{equation}
with the normalization condition
\begin{equation}
|\alpha_a|^2+|\beta_a|^2+|\gamma_a|^2+|\delta_a|^2=1  
\end{equation} and a similar parametrization for $\ket{B}_E, \ket{C}_E, \ket{D}_E$, with indices $b, c, d$, respectively.
A partial-trace operation on $\rho^{(1)}_{ABE}$ over Eve's part leads to the state $\rho^{(1)}_{AB}$, which corresponds to the state shared by Alice and Bob after Eve's unitary interaction. It has been shown in \cite{Kra05,Ren05a} that the BB$84$ and the six-state protocol permit to characterize the state $\rho^{(1)}_{AB}$ as Bell-diagonal, parametrized by the quantum bit error rate $Q$.
\begin{eqnarray}
\hspace*{-0.5cm}
 \rho^{(1)}_{AB}&=&\lambda_1\ket{\Psi^+}\bra{\Psi^+}+\lambda_2\ket{\Psi^-}\bra{\Psi^-} 
+\lambda_3\ket{\Phi^+}\bra{\Phi^+}+\lambda_4\ket{\Phi^-}\bra{\Phi^-},
\end{eqnarray} with the Bell-states
\begin{eqnarray}
 \ket{\Psi^\pm}&=&\frac{1}{\sqrt{2}}\left(\ket{01}\pm\ket{10}\right) \\
 \ket{\Phi^\pm}&=&\frac{1}{\sqrt{2}}\left(\ket{00}\pm\ket{11}\right),
\end{eqnarray}
and the parameters
\begin{equation}
 \lambda_1=1-\frac{3}{2}Q, \lambda_2=\lambda_3=\lambda_4=\frac{Q}{2}
\end{equation} for the six-state protocol, while
\begin{equation}
 \lambda_1=1-2Q+\lambda_4, \lambda_2=\lambda_3=Q-\lambda_4, \lambda_4 \in \left[0,Q\right]
\end{equation} for the BB$84$ protocol. Note that the symmetry properties of
$\rho^{(1)}_{AB}$
are preserved by adding  symmetric depolarizing noise (see Definition~\ref{defi:deponoise}).

We can express the $QBER$ $Q$ as a function of the noise parameter $p_a$, which is introduced by the depolarizing channel,
and Eve's disturbance $D$: 
\begin{equation}
\label{noiseadds}
 Q=(1-p_a)D+\frac{p_a}{2}.
\end{equation} 
For the six-state protocol
 we obtain the following additional conditions on Eve's probes:
\begin{equation}\label{EQ:cond0}
 \bracket{A}{B}_E=\bracket{A}{D}_E=\bracket{B}{C}_E=\bracket{D}{C}_E=\bracket{B}{D}_E=0
\end{equation} and 
\begin{equation}\label{EQ:cond}
 \bracket{A}{C}_E=\frac{1-2Q}{(1-p_a)(1-D)}.
\end{equation}
W.l.o.g. we can choose $\ket{D}_E=\ket{00}$ and $\ket{B}_E=\ket{11}$. It follows that
\begin{eqnarray}
 \ket{A}_E&=&\beta_a \ket{01}+\sqrt{1-|\beta_a|^2}\ket{10}, \\
 \ket{C}_E&=&\beta_c \ket{01}+\sqrt{1-|\beta_c|^2}\ket{10}.
\end{eqnarray}
By using \Eref{EQ:cond} we eliminate $\beta_a$, such that with the constraints in \Eref{EQ:cond0} 
the state $\rho^{(1)}_{ABE}$ contains only one unknown parameter, namely $\beta_c$.

Analogously, we get for the BB$84$ protocol the following constraints:
\begin{equation}
 \bracket{A}{B}_E=\bracket{A}{D}_E=\bracket{B}{C}_E=\bracket{D}{C}_E=0
\end{equation} and 
\begin{eqnarray}
 \bracket{B}{D}_E &=& \frac{Q-2\lambda_4}{(1-p_a)D} \label{EQ:cond1}, \\
 \bracket{A}{C}_E &=& \frac{1-3Q+2\lambda_4}{(1-p_a)(1-D)}.\label{EQ:cond2}
\end{eqnarray}
W.l.o.g. we can choose $\ket{A}_E=\ket{11}$ and $\ket{B}_E=\ket{00}$. It follows that
\begin{eqnarray}
 \ket{C}_E &=& \sqrt{1-|\delta_c|^2}\ket{10}+\delta_c\ket{11}, \\
 \ket{D}_E &=& \alpha_d \ket{01}+\sqrt{1-|\alpha_d|^2}\ket{10}.
\end{eqnarray}
By using \Eref{EQ:cond1} and \Eref{EQ:cond2} we can reduce the unknown parameters of the state
$\rho^{(1)}_{ABE}$ to the single parameter $\lambda_4$.

Remember that $\rho^{(1)}_{ABE}$ describes the quantum state shared by Alice, Bob and Eve after the state distribution step for noise scenario $1$. Let us denote the classical-classical-quantum state that results from local von Neumann measurements performed by Alice and Bob by $\rho_{XYE}$. The states $\rho_{XE}$ and $\rho_{XY}$, which are needed for the calculation of the asymptotic key rate in \Eref{EQ:asymrate}, follow directly by a partial-trace operation on Bob's part and Eve's part, respectively. 
 
The unknown parameter $\beta_c$ ($\lambda_4$) for the six-state protocol (BB$84$ protocol) has to be chosen in such a way that it minimizes the asymptotic key rate in \Eref{EQ:asymrate}, such that these states realize Eve's best strategy.
After including the finite-key corrections (\Eref{EQ:finitekey}) into the optimization, the key rate is now fully determined by the noise parameter $p_a$ and the disturbance $D$, such that the effects of noise can be calculated, also in the regime of a finite number of signals.

\section{Results}\label{SEC:results}
In this section we present
our results on the secret key rate in the noisy scenario described above. We will discuss
 two possible interpretations of our results: In subsection \ref{delinoise} we consider the case that the noise is introduced deliberately
by Alice. In subsection \ref{channelnoise} we analyse the case where the noise is given by the channel and did not originate from the eavesdropper. The finite-key rate $r$ (\Eref{EQ:finitekey}) will be calculated in both cases for a total security parameter of $\eps{}=10^{-9}$. The results are obtained from a numerical optimization procedure, which maximizes the key rate with respect to the parameters $m,\bar{\varepsilon},\eps{PE},\eps{EC},\eps{PA}$, while minimizing with respect to the parameters $\beta_c$ ($\lambda_4$), for the six-state (BB$84$) protocol.
 
\subsection{Introducing noise deliberately}
\label{delinoise}
In the following we illustrate the effect of deliberately added noise (see \Sref{SEC:noisescenarios}) on the finite key rate $r$ (see \Eref{EQ:finitekey}).

\begin{figure}
\centerline{\includegraphics[width=0.5\textwidth]{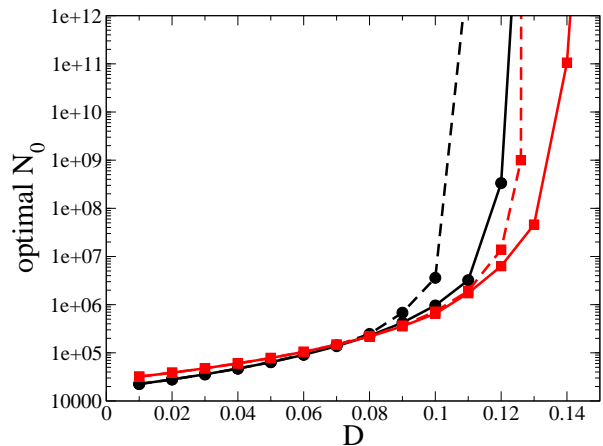}}
 \caption{Comparison of the optimal minimal number $N_0$ to extract a non-zero key ($\eps{}=10^{-9}$) versus the QBER $D$ introduced by Eve for the BB$84$ (circles (black)) and the six-state protocol (squares (red)); straight line: noise scenario $1$ (see \Sref{SEC:noisescenarios}), dashed line: no noise.}
 \label{FIG:asymratemergefinite}
 \end{figure}

\begin{figure}
\centerline{\includegraphics[width=0.5\textwidth]{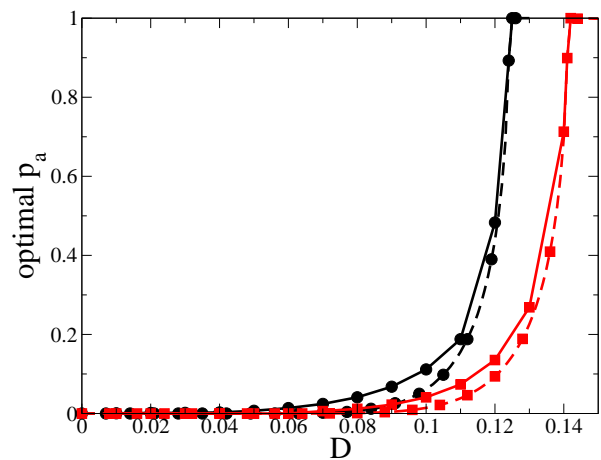}}
\caption{Comparison of the optimal noise parameter that minimizes $N_0$ (straight lines) for the finite-key rate (\Eref{EQ:finitekey}) and the one that maximizes the asymptotic key rate (\Eref{EQ:asymrate}) (dashed lines) versus the QBER $D$ introduced by Eve for noise scenario $1$ (see \Sref{SEC:noisescenarios}); circles (black): BB$84$, squares (red): six-state.}
\label{FIG:noiseoptfinite}
\end{figure}

In \Fref{FIG:asymratemergefinite} the behaviour of $N_0$, the minimal number of signals that is needed to extract a non-zero key, with respect to the disturbance $D$ is shown for an optimal noise parameter $p_a$ (see \Fref{FIG:noiseoptfinite}) for the BB$84$ and the six-state protocol. In comparison to the noiseless case we obtain a beneficial effect on the finite key rate by introducing quantum noise $p_a$:
In the six-state (BB$84$) protocol with $D=0.12$ ($D=0.1$) the improvement in the minimal number of signals $N_0$ is of the 
 order of a million signals. Additionally, we find that noise enables us to extract a non-zero key for higher disturbances than in the noiseless case: We recover for our case of a finite number of signals the result of  \cite{Kra05}, which states that the maximum tolerated error rate introduced by Eve to extract a non-zero key is shifted from $12.6 \%$ to $14.1 \%$ ($11.0 \%$ to $12.4 \%$) for the six-state (BB$84$) protocol, in the asymptotic limit $N_0 \rightarrow \infty$.

In \Fref{FIG:noiseoptfinite} we show the optimal noise parameter $p_a$ that minimizes $N_0$ and compare it to the optimal noise parameter $p_a$ that maximizes the asymptotic key rate (\Eref{EQ:asymrate}) for various disturbances $D$ for the BB$84$ and the six-state protocol. 
It turns out that the optimal noise parameter $p_a$ for the finite case is always higher than the one for the asymptotic
case. In particular, for the asymptotic key rate 
 the optimal $p_a$ becomes non-zero for around $D=0.096$ ($D=0.083$) for the six-state (BB$84$) protocol, 
while the benefit for the threshold $N_0$ 
in the finite scenario
appears already for disturbances around $0.08$ ($0.06$) for the six-state (BB$84$) protocol. 

\begin{figure}
\centerline{\includegraphics[width=0.5\textwidth]{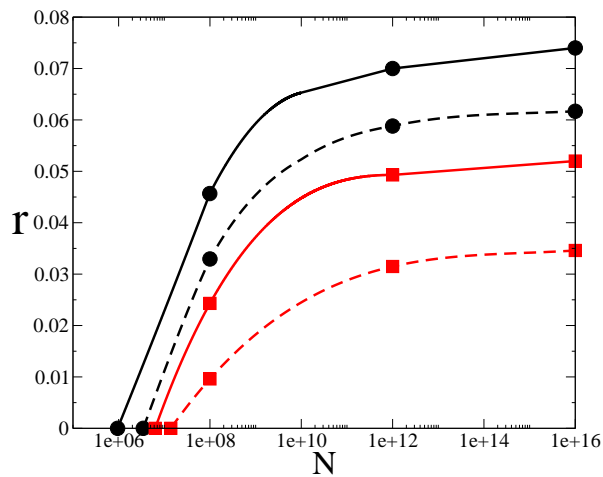}}
\caption{Comparison of the finite-key rate (\Eref{EQ:finitekey}) versus signals $N$ for a fixed disturbance $D$ for the BB$84$ protocol ($D=0.1$) (circles (black)) and the six-state protocol ($D=0.12$) (squares (red)); straight line: noise scenario $1$ (see \Sref{SEC:noisescenarios}) for optimal noise parameter, dashed line: no noise. Lines are drawn to guide the eye.}
\label{FIG:noiseoptfiniteoverall}
\end{figure}

In \Fref{FIG:noiseoptfiniteoverall} we show the optimal secret key rate $r$ as a function of the 
number of signals $N$ for a fixed disturbance $D=0.1$ for the BB$84$ protocol and $D=0.12$ for the six-state protocol, and compare it to the case without added noise. We obtain that the effect of noise on the finite-key rate is more beneficial than the effect on the asymptotic key rate,
when taking the relative increase of the key rate as figure of merit. For example, for $N=10^8$ signals we have an increase of $39\%$ ($153\%$) in the key rate, whereas the benefit for $N=10^{16}$ is only about $20\%$ ($50\%$) for the BB$84$ (six-state) protocol. 

\subsection{Noise given by the channel}
\label{channelnoise}
\begin{figure}[h]
\centerline{\includegraphics[width=0.5\textwidth]{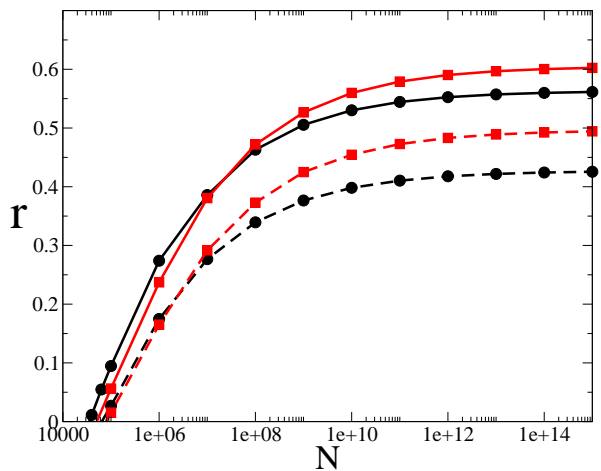}}
\caption{Comparison of the finite-key rates (\Eref{EQ:finitekey}) ($\eps{}=10^{-9}$) versus 
number of signals $N$ for various noise parameters $p_b$ with $QBER=5\%$ for the BB$84$ (circles (black)) and the six-state protocol (squares (red)); dashed lines: $p_b=0$, straight lines: $p_b=5\%$.}
\label{FIG:finiterate}
\end{figure}

The equivalence of noise scenario $1$ and $2$ for $p_a=p_b$ allows us to interpret the results obtained in \Sref{SEC:keyrate} in another way. In contrast to adding noise deliberately the number $p_b$ can be interpreted as the amount of noise that is introduced by the used quantum channel, which is unavoidable in real QKD settings, and not necessarily dedicated to the eavesdropper. This interpretation describes the situation in real experiments, where the assumption of unconditional security, i.e. all errors introduced by the channel have to be attributed to the eavesdropper, is over-pessimistic \cite{Qi07,Sca09}. 
If one makes the realistic assumption that Eve cannot replace the noisy channel by a noisefree one, the channel noise does not
lead to knowledge of Eve about the key, and the key rate will thus increase.
In \Fref{FIG:finiterate} the finite-key rate (\Eref{EQ:finitekey}) is shown as a function of the 
number of signals $N$ sent through the channel for a fixed $QBER$ ($Q=5\%$)
for different values of the noise parameter $p_b$ for the six-state and BB$84$ protocol. The measured $QBER$ contains both  the
 noise $p_b$ that we attribute to the channel and the noise $D$ that is related to Eve's unitary interaction. 
For the explicit connection between these different types of noise see Eq. (\ref{noiseadds}).
Taking this fact into account leads to remarkably higher key rates, as shown in \Fref{FIG:finiterate}. For example, for $N=10^8$ signals, without added noise the key rate in the six-state (BB$84$) protocol is $0.37$ ($0.34$), while for channel noise of $p_b=0.05$ it is $0.47$ ($0.46$) for the six-state (BB$84$) protocol. 

\section{Conclusions}\label{SEC:conclusion}
In this article we have shown that the presence
of quantum noise can improve secret key rates, in particular in the realistic
scenario of a finite number of resources.
We have investigated the effect of different noise scenarios on an $\eps{}$-secure key rate for the BB$84$ and the six-state protocol in the entanglement-based scheme, for a finite number of signals. Our results can be interpreted in two  ways:
First, when taking the
 view that noise is added deliberately, it turns out that the effect of adding depolarizing noise to the state (before the state transmisson) 
is equal to the benefit gained by adding classical noise, i.e. when Alice performs probabilistic bit-flips on her measured bit string. 
We obtain that 
for both the BB$84$ and the six-state protocol
the benefit (concerning the key rate)
of adding noise is higher in the regime of a finite number of signals than for the asymptotic key rate. 
Second, under the realistic assumption that a channel itself introduces noise unavoidably, i.e.
the noise  is not necessarily created by the eavesdropper, the secret key rate increases significantly
with respect to the ''worst case'', where all noise is attributed to Eve's intervention.
 This improvement comes from the fact that the errors from the quantum channel 
do not give Eve information about the key.  This approach avoids the over-pessimistic assumption of unconditional security,
and is thus meaningful for realistic experiments.

\ack
We would like to thank Silvestre Abruzzo and Sylvia Bratzik for helpful discussions. This work was financially supported by  Deutsche Forschungsgemeinschaft (DFG) and  Bundesministerium f\"ur Bildung und Forschung (BMBF) project QuOReP.

\section*{References}

\end{document}